\documentclass[10pt]{iopart}
\usepackage{graphicx}
\usepackage[dvips]{color}

\newcommand {\be}{\begin{equation}}
\newcommand {\ee}{\end{equation}}
\newcommand {\bea}{\begin{eqnarray}}
\newcommand {\eea}{\end{eqnarray}}

\begin{document}

\title{Macroscopic quantum tunnelling of Bose-Einstein condensates in a
finite potential well}
\author{L.\ D.\ Carr$^{1}$, M. J. Holland$^{1}$, and B. A. Malomed$^{2}$}

\begin{abstract}
Bose-Einstein condensates are studied in a potential of finite
depth which supports both bound and quasi-bound states. This
potential, which is harmonic for small radii and decays as a
Gaussian for large radii, models experimentally relevant optical
traps. The nonlinearity, which is proportional to both the number
of atoms and the interaction strength, can transform bound states
into quasi-bound ones. The latter have a finite lifetime due to
tunnelling through the barriers at the borders of the well. We
predict the lifetime and stability properties for repulsive and
attractive condensates in one, two, and three dimensions, for both
the ground state and excited soliton and vortex states. We show,
via a combination of the variational and WKB approximations, that
macroscopic quantum tunnelling in such systems can be observed on
time scales of 10 milliseconds to 10 seconds.
\end{abstract}

\maketitle

\address{
$^1$ JILA, University of Colorado and National Institute of Standards and Technology, Boulder, CO 80309-0440\\
$^2$ Department of Interdisciplinary Studies, Faculty of
Engineering, University of Tel Aviv, Israel}

\section{Introduction}

\label{sec:intro}

The tunnelling of a particle through a potential barrier is a
fundamental effect in quantum mechanics~\cite{landau1977}.
\emph{Macroscopic quantum tunnelling} is the tunnelling of a
many-body wavefunction through a potential barrier, and therefore
provides a more stringent test of the validity of quantum
mechanics than the one particle case~\cite{leggett2001}. One place
where the study of macroscopic quantum tunnelling is particularly
experimentally accessible is in the tunnelling of a Bose-Einstein
condensate (BEC) out of an optical trap. Recently, we showed that
optical traps, which are of finite depth, can support both bound
and quasi-bound states, and that the nonlinearity of the BEC mean
field can be used to tune between
them~\cite{carr2004b,witthaut2005,adhikari2005}. In this article,
we calculate the lifetime of such quasi-bound states for both the
ground state and excited soliton and vortex states.  This provides
a straightforward experimental observable for the occurrence of
macroscopic quantum tunnelling of a BEC.

BEC's exhibit different kinds of tunnelling phenomena. We consider
the most direct generalization of the single-particle
case~\cite{landau1977}, the tunnelling of the mean field through a
barrier via the Gross-Pitaevskii, or nonlinear Schr\"{o}dinger
equation (NLS)~\cite{dalfovo1999}. We emphasize that this is a
\emph{nonlinear} tunnelling problem in the mean-field
approximation~\cite{carr2002c}. Experimentally, tunnelling of the
mean field in a double-well~\cite{shin2004a,shin2004c} or
sinusoidal lattice potential~\cite{anderson1998} in configuration
space, as well as in spin space for multiple-spin-component
BEC's~\cite{stamperkurn1999}, has been investigated. There have
been various theoretical studies related to these
experiments~\cite{kasamatsu2001,liu2002,pu2002,lee2003,huepe2003,sakellari2004,shchesnovich2004,shchesnovich2005}.
This is to be contrasted with tunnelling of the whole condensate
in a variational parameter space, as was previously considered in
the context of the collapse of a metastable attractive BEC in
three
dimensions~\cite{kagan1996,shuryak1996,stoof1997,ueda1998,sackett1998}
and in quantum evaporation of a bright soliton in an expulsive
harmonic trap~\cite{carr2002c}.

In this work, we consider a spherically symmetric trap in the form
of a parabolic potential times a Gaussian, where the width of the
Gaussian envelope is much greater than the harmonic oscillator
length. This models the optical trap used in many experiments on
BEC's (see, for instance,~\cite{barrett2001}). A potential offset
$V_{0}$ at the origin, which is important~\cite{carr2004b} in
preventing collapse of attractive condensates in three
dimensions~\cite{sulem1999}, can be added with an additional blue-
or red-detuned laser beam focused at the center of the trap. We
take advantage of semi-classical methods and employ a
variational-WKB formalism to calculate the lifetime of the
condensate held in such a potential~\cite{carr2002c}. In addition,
we calculate three critical values of the nonlinearity which could
be observed in experiments: the point at which the condensate
collapses, termed the \emph{collapse nonlinearity}; the point at
which the quasi-bound states are transformed into bound states,
called the \emph{critical nonlinearity}; and the point at which
the condensate is pushed out over the top of the potential and
quasi-bound states cease to exist at all, called the \emph{maximum
nonlinearity}. The collapse nonlinearity is observable as the
point where the condensate first implodes.  The critical
nonlinearity is observable as the point where the lifetime first
becomes longer than inverse experimental loss rates.  The maximum
nonlinearity is observable as the point where the condensate
commences to expand.

We emphasize that even at the level of the mean field
approximation, there are features that are distinctly different
from the tunnelling of a single particle. As the tunnelling rate
depends on the number of atoms remaining in the well, the lifetime
is not simply inversely proportional to the rate as in the linear
Schr\"{o}dinger equation, but must be determined by an integral
over each step of the tunnelling. The final state of the
tunnelling process need not be that of zero probability of there
being an atom remaining in the well. Instead, a quasi-bound
repulsive condensate can decay towards a final bound state in
which a finite number of atoms remain in the well. A quasi-bound
attractive condensate can decay towards an unbound state so that
the atoms spill out over the top of the well.

In Sec.~\ref{sec:setup}, our application of the variational-WKB
method to this nonlinear tunnelling problem is explained in
detail. In Sec.~\ref{sec:ground} the lifetime and the three
above-mentioned critical points are studied for the ground state
in one, two, and three dimensions. Excited states are treated in
Sec.~\ref{sec:soliton} in one dimension for dark solitons and
bright twisted solitons, and in Sec.~\ref{sec:vortex} for vortices
in two dimensions. In all cases we consider both repulsive and
attractive condensates.

\section{Fundamental equations and methods}

\label{sec:setup}

The time-independent isotropic NLS with an external potential of the form
described in Sec.~\ref{sec:intro} may be written as
\begin{eqnarray}
-\frac{\hbar ^{2}}{2M}\nabla
^{2}\tilde{\Psi}+\tilde{V}(\tilde{r})\tilde{\Psi}+\tilde{U}_{D}|\tilde{\Psi}|^{2}\tilde{\Psi}
=\tilde{\mu}\tilde{\Psi}\,,
\\
\tilde{V}(\tilde{r}) =\left( \tilde{V}_{0}+\frac{1}{2}M\omega
^{2}\tilde{r}^{2}\right) \exp \left( -\tilde{r}^{2}/2\ell
_{\mathrm{Gauss}}^{2}\right) \,,
\\
\beta _{D}\int_{0}^{\infty
}d\tilde{r}\,\tilde{r}^{D-1}|\tilde{\Psi}|^{2} =1\,,
\end{eqnarray}where the tildes indicate that the respective variables and
parameters are measured in physical units. Here $\tilde{\Psi}$ is
the mean-field wavefunction with the integral norm set to unity,
$M$ is the atomic mass, $\tilde{\mu}$ is a complex eigenvalue
which we term the chemical potential, $\ell _{\mathrm{Gauss}}$ is
the width of the Gaussian envelope, $\omega $ is the angular
frequency of the parabolic trap, and $\tilde{V}_{0}$ is the
potential offset at the origin. The constant
\begin{equation}
\beta _{1,2,3}\equiv 2,2\pi ,4\pi
\end{equation}for the spatial dimension $D=1,2,3$. The coupling constants are renormalized
appropriately for the dimensionality:
\begin{equation}
\tilde{U}_{1,2,3}=(2\hbar \omega _{\perp })aN\,,\,\left(
\sqrt{\frac{8\pi \hbar ^{3}\omega _{z}}{M}}\right) \,aN\,,\,\left(
\frac{4\pi \hbar ^{2}}{M}\right) aN\,\,,
\end{equation}where $a$ is the $s$-wave scattering length and $N$ is the number of atoms.
The transverse oscillator frequencies $\omega _{\perp },\omega _{z}$ must be
sufficiently high so as to reduce the effective dimensionality of the mean
field of the BEC~\cite{carr2000e,olshanii1998,petrov2000,petrov2000b}.

The NLS can be conveniently rescaled to dimensionless form:
\begin{eqnarray}
-\frac{1}{2}\nabla^2\Psi + V(r)\Psi +U_{D}|\Psi |^{2}\Psi = \mu \Psi \,,
\label{eqn:GPE} \\
V(r) = \left( V_{0}+\frac{1}{2}r^{2}\right) \exp \left( -\alpha
\,r^{2}\right)\,,  \label{eqn:pot} \\
\beta _{D}\int_{0}^{\infty }dr\,r^{D-1}|\Psi |^{2} = 1 \,,  \label{eqn:norm}
\end{eqnarray}
where all variables are in units of the harmonic oscillator energy
$\hbar\omega $ and length $\ell_{\mathrm{ho}}\equiv \sqrt{\hbar
/m\omega }$. The coupling constants become
\begin{equation}
U_{1,2,3}=a N/\ell_{{\mathrm{ho}}}\, , \sqrt{2/\pi}\,a
N/\ell_{{\mathrm{ho}}}\, , a N/\ell_{{\mathrm{ho}}}\,.
\end{equation}
The parameters $V_0$ and
\begin{equation}
\alpha \equiv (\ell _{\mathrm{ho}}/\ell_{\mathrm{Gauss}})^{2}
\end{equation}
characterize the structure of the potential. For a broad Gaussian envelope,
which pertains to the experimentally available optical trap described in
Sec.~\ref{sec:intro}, $\alpha \ll 1$.

Approximate solutions to Eq.~(\ref{eqn:GPE}) with potential
(\ref{eqn:pot}) can be obtained via the variational method. The
ansatz we use in each of the cases to be considered below contains
a Gaussian factor of form $A\,\exp [-r^{2}/(2\rho ^{2})]$, with
the amplitude $A$ and the width $\rho $ taken as variational
parameters. Substituting the ansatz into the normalization
condition (\ref{eqn:norm}) and the Lagrangian of the NLS (for the
time being, $\mathrm{Im}(\mu )$ is disregarded), one minimizes the
Lagrangian with respect to $\rho $ and $A$.  Then one obtains a
system of equations for the nonlinearity $U_{D}$ and the real part
of the chemical potential $\mathrm{Re}(\mu )$ in terms of given
parameters of the system, $D$, $\alpha $, and $V_{0}$. The
solution is stable in the framework of the time-dependent radially
symmetric NLS if
\begin{equation}
d\mathrm{Re}(\mu )/d|U_{D}|\leq 0\,,  \label{eqn:vk}
\end{equation}which is known as the Vakhitov-Kolokolov (VK) criterion~\cite{vakhitov1973}.
An important point is that our choice of ansatz truncates the
solution space by requiring the wavefunction decay as
$r\rightarrow \infty $, eventually enabling us to find quasi-bound
states in the form of eigenstates with complex eigenvalues.

To find the imaginary part of the chemical potential one can
separately apply the WKB
approximation~\cite{landau1977,brack1997}. The tunnelling rate
$\gamma$ in $D$ dimensions is given by the standard expressions
\begin{eqnarray}
\gamma  &=&\nu \,\exp \left( -2\int_{r_{1}}^{r_{2}}dr\,|p(r)|\right) \,,
\label{eqn:rate} \\
p(r) &\equiv &\sqrt{2[\mu -V_{D}^{\mathrm{eff}}(r)]}\,, \\
\nu ^{-1} &\equiv &4\int_{0}^{r_{1}}\frac{dr}{|p(r)|}\,,
\end{eqnarray}where the endpoints $r=r_{1}$ and $r=r_{2}$ are found from setting the
semiclassical momentum $p(r)=0$ and
\begin{eqnarray}
V_{1}^{\mathrm{eff}}(r) &\equiv &V(r)+U_{1}|\phi _{1}|^{2}\,, \label{eqn:effpot1}\\
V_{2}^{\mathrm{eff}}(r) &\equiv
&V(r)-\frac{1}{8r^{2}}+\frac{U_{2}}{r}|\phi
_{2}|^{2}\,, \label{eqn:effpot2}\\
V_{3}^{\mathrm{eff}}(r) &\equiv &V(r)+\frac{U_{3}}{r^{2}}|\phi
_{3}|^{2}\,.\label{eqn:effpot3}
\end{eqnarray} Here $\nu $ is the semiclassical oscillation frequency
in the well. The imaginary part of the chemical potential is then
given by
\begin{equation}
\mathrm{Im}(\mu )=-\gamma /2\, .
\end{equation} Note that $\gamma $ must be multiplied by $2$ in the 1D
case, to account for tunnelling from both sides of the well. The
transformed wavefunctions in
Eqs.~(\ref{eqn:effpot1})-(\ref{eqn:effpot3}) are given by the
standard expressions
\begin{equation}
\phi _{1,2,3}=\Psi \,,\,\sqrt{r}\Psi \,,\,r\Psi \,.
\end{equation}

In the case of axially symmetric excited states in 2D, i.e., vortices, it is
useful to make the phase winding number explicit:
\begin{equation}
\phi _{2}=\phi _{2,m}e^{im\theta }\,.  \label{eqn:2dtransform}
\end{equation}Then the transformed wavefunction $\phi _{2,m}$ leads to the simplified NLS
\begin{equation}
-\frac{1}{2}\frac{\partial ^{2}}{\partial r^{2}}\phi
_{2,m}+V_{2,m}^{eff}\phi _{2,m}=\mu \,\phi _{2,m}\,,  \label{eqn:gpe2d}
\end{equation} where the effective potential is
\begin{equation}
V_{2,m}^{eff}=V(r)+\frac{m^{2}-1/4}{2\,r^{2}}+\frac{U_{2}}{r}|\phi
_{2,m}|^{2}\,.  \label{eqn:effpot2d}
\end{equation}Note that the centrifugal barrier changes sign for $m\neq 0$, so that the
semiclassical period of oscillation in the well must be redefined as
\begin{equation}
\nu ^{-1}=2\int_{r_{0}}^{r_{1}}\frac{dr}{|p(r)|}\,,
\end{equation}where $r_{0}$ is the inner turning point created by the centrifugal
barrier. The outer potential barrier is delimited by $r_{1},\,r_{2}$.

The main experimental observable we consider is the lifetime of
the trapped condensate. This is not given by $1/\gamma $, as in a
linear system, but must be found from the rate equation
\begin{equation}
dN/dt=-\gamma (N)N\,,
\end{equation}where $\gamma (N)$ is the nonlinear tunnelling rate and $N$ is the number of
atoms in the BEC remaining in the trap, i.e., between the classical turning
points. This leads to the integral
\begin{equation}
T=\int_{N_{0}}^{N_{0}/e}dN/[-\gamma (N)N]\,,  \label{eqn:lifetime}
\end{equation}
where $T$ is the lifetime and $N_{0}$ is the initial number of
atoms in the well. Here we have assumed that the atomic
interaction strength $a$ and the harmonic oscillator length $\ell
_{\mathrm{ho}}$ are constant in time. Therefore, in practice, we
integrate over the norm $U_{D}$, which is proportional to $N$.

Besides the lifetime, there are three important quantities we will
calculate. The maximum nonlinearity $U_{D}^{\mathrm{max}}$ is the
point at which the condensate spills out over the top of the well
and a quasi-bound state ceases to exist. The critical nonlinearity
$U_{D}^{\mathrm{crit}}$ is the point at which the real part of the
chemical potential changes from positive to negative, the
imaginary part goes to zero, and the quasi-bound state is
transformed into a genuine bound state. The collapse nonlinearity
$U_{D}^{\mathrm{coll}}$ is the point at which the condensate
collapses in two and three dimensions, as we will explain in more
detail in the following section.

There is an important subtlety in the limits of integration in
Eq.~(\ref{eqn:lifetime}). We can define an
$N^{\mathrm{crit}}\propto U_{D}^{\mathrm{crit}}$ and
$N^{\mathrm{max}}\propto U_{D}^{\mathrm{max}}$. If
$U_{D}^{\mathrm{crit}}>0$, then the nonlinearity is always
repulsive for tunnelling processes. The upper limit of
Eq.~(\ref{eqn:lifetime}) must be replaced with
$(N_{0}-N_{\mathrm{crit}})/e+N_{\mathrm{crit}}$, and we define the
lifetime with respect to decay to a stable state. That is to say,
no more than $(N_{0}-N_{\mathrm{crit}})$ particles will ever leave
the well. On the other hand, if $U_{D}^{\mathrm{max}}<0$, then the
nonlinearity is always attractive. The upper limit of
Eq.~(\ref{eqn:lifetime}) must then be replaced with
$(N_{0}-N_{\mathrm{max}})/e+N_{\mathrm{max}}$, and we define the
lifetime with respect to decay to a non-stationary state. In this
case, after $(N_{0}-N_{\mathrm{max}})/e$ particles have left the
well via quantum mechanical tunnelling, the rest simply flow over
the top classically. These cases are extremely different from
tunnelling of a single atom in the linear Schr\"{o}dinger
equation, where the final state always corresponds to zero atoms
remaining in the well.

\section{Tunnelling of the Ground State in One, Two, and Three Dimensions}

\label{sec:ground}

Taking the variational ansatz as a Gaussian,
\begin{equation}
\Psi _{\mathrm{gs}}=A\,\exp \left( -\frac{r^{2}}{2\rho ^{2}}\right) \,,
\label{eqn:ansatz}
\end{equation} one minimizes the Lagrangian corresponding to Eq.~(\ref{eqn:GPE}).
The result is a system of Euler-Lagrange equations for the
chemical potential and
nonlinearity~\cite{sakurai1994,malomed2002}. These take the form
\begin{eqnarray}
\mathrm{Re}(\mu ) &=&\frac{1}{4}\left\{ (-4+D)\rho ^{-2}+(1+\alpha \rho
^{2})^{-2-D/2}\right.   \nonumber \\
&&\left. \times \left[ 4V_{0}+(4+D)\rho ^{2}\right. \left. -\alpha
(D+4V_{0}\alpha )\rho ^{4}\right] \right\} \,,\,\,\,\,\,\,  \label{eqn:mu1}
\end{eqnarray}\begin{eqnarray}
U_{D} &=&\Gamma (D/2)2^{-2+D/2}\beta \rho ^{-2+D}(1+\alpha \rho
^{2})^{-2-D/2}  \nonumber \\
&&\times \left\{ -2(1+\alpha \rho ^{2})^{2+D/2}+\rho ^{4}[2-4V_{0}\alpha
-\alpha (D+4V_{0}\alpha )\rho ^{2}]\right\} ,::::::  \label{eqn:U01}
\end{eqnarray}where $\Gamma (D/2)$ is a Gamma function and $D$ is the dimensionality.

\begin{figure}[t]
\centerline{
\includegraphics[width=3in,angle=0,keepaspectratio]{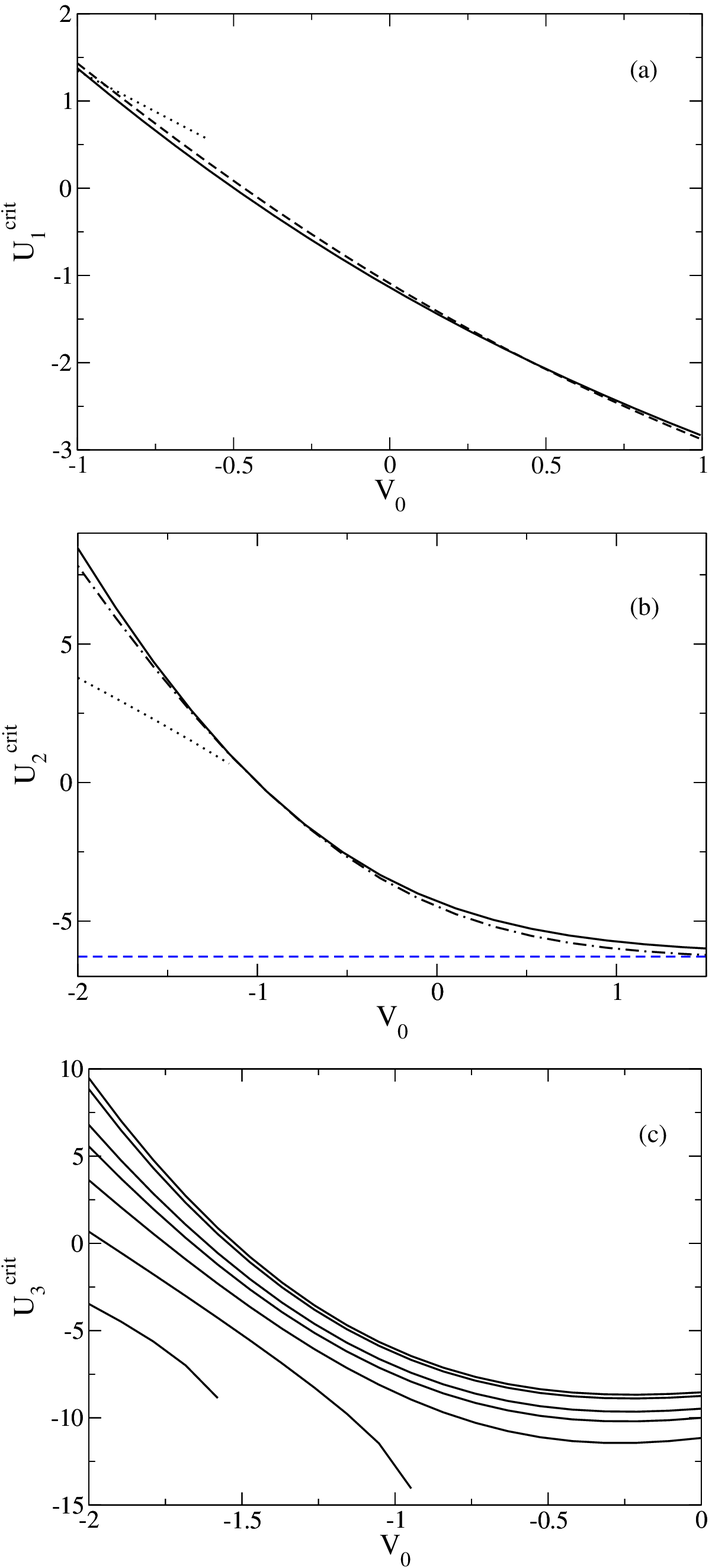}}
\caption{\textit{Ground state transition.} Shown is the critical
nonlinearity for a transition from a bound to a quasi-bound state,
as a function of the potential offset $V_{0}$. (a) One dimension,
$\protect\alpha =1,1/2,1/10$ (dotted, dashed, solid curves), where
$\protect\alpha \equiv \ell _{\mathrm{ho}}^{2}/\ell
_{\mathrm{Gauss}}^{2}$. There is no collapse in one dimension. (b)
Two dimensions: $\protect\alpha =1,1/2,1/10$ (dotted, dot-dashed,
solid curves); the critical nonlinearity for collapse,
$U_{2}^{\mathrm{coll}}=-2\protect\pi $ is indicated by the blue
dashed curve. Note that the condition
$U_{2}^{\mathrm{crit}}>U_{2}^{\mathrm{coll}}$ always holds. (c)
Three dimensions: $\protect\alpha =1,1/2,1/3,1/4,1/5,1/6,1/7$ from
bottom left to top right. The collapse point is described in
Fig.~\protect\ref{fig:2}.} \label{fig:1}
\end{figure}

The transition from a quasi-bound to a bound state occurs when the
real part of the chemical potential changes sign from positive to
negative, since our potential $V(r)\rightarrow 0^{+}$ as
$r\rightarrow \infty $. The nonlinearity at which this occurs we
termed $U_{D}^{\mathrm{crit}}$. In Fig.~\ref{fig:1} is shown the
dependence of $U_{D}^{\mathrm{crit}}$ on the potential offset
$V_{0}$, as defined in Eq.~(\ref{eqn:pot}). The cases of $D=1,2,3$
dimensions are illustrated in separate panels in the figure. The
different curves in each panel pertain to different values of the
trap-shape parameter $\alpha $. These two continuous parameters,
$V_{0}$ and $\alpha $, and the discrete one, $D$, are the only
given constants in the system. The potential offset $V_{0}$ can be
controlled by a red- or blue-detuned laser focused at the center
of the trap, as previously mentioned.

\begin{figure}[t]
\centerline{
\includegraphics[width=3in,angle=0,keepaspectratio]{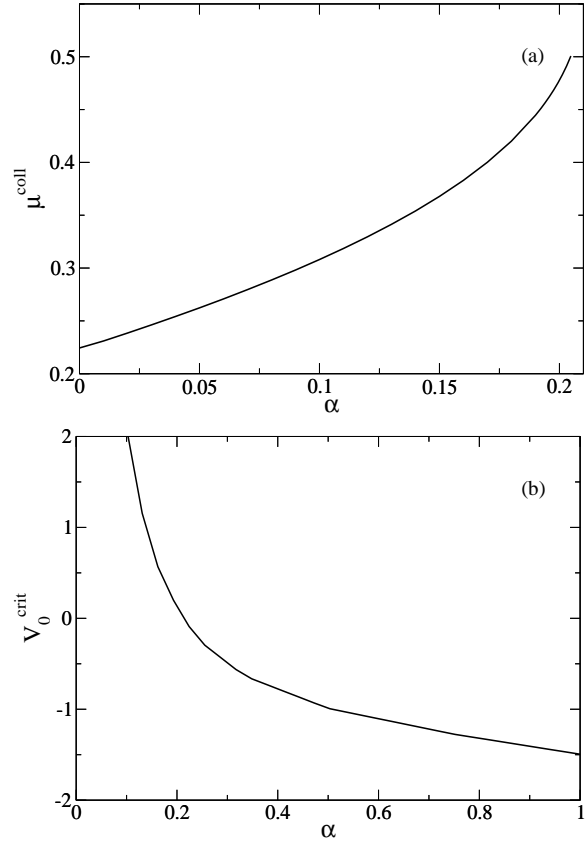}}
\caption{\textit{Ground state parameters, three dimensions.} (a)
Shown is the dependence of the collapse point on the trap-shape
parameter $\protect\alpha $ for zero offset ($V_{0}=0$).  For
nonzero $V_{0}$ and $\protect\alpha \ll 1$, $\protect\mu
^{\mathrm{coll}}\simeq \protect\mu
^{\mathrm{coll}}(V_{0}=0)+V_{0}$. In order to obtain an observable
transition from a quasi-bound to a bound state, it is necessary
that $\protect\mu ^{\mathrm{coll}}<0$. (b) Shown is the critical
offset $V_{0}^{\mathrm{crit}}$ needed to obtain \textit{any}
quasi-bound state for a given $\protect\alpha $: $V_{0}\leq
V_{0}^{\mathrm{max}}$. } \label{fig:2}
\end{figure}

However, the existence of the transition from a quasi-bound to a
bound state does not mean that the ground state solution we
obtained is stable. When the nonlinearity is negative, so that the
atoms attract each other, the condensate can collapse. In three
dimensions with $V(r)=0$, any initial condition leads to
collapse~\cite{sulem1999}.  As is well-known for BEC's, the
imposition of a nonzero potential $V(r)$ can create a metastable
solution with a lifetime much longer than that of the BEC, so that
it is experimentally stable. This requires a nonlinearity which is
not too strongly negative, $U_{3}>U_{3}^{\mathrm{coll}}$, where
$U_{3}^{\mathrm{coll}}$ is the critical point for collapse. This
critical point can be determined variationally or
numerically~\cite{ruprecht1995,perez1997,carr2004b}. In order to
observe the transition from a quasi-bound to a bound state, it is
necessary that $U_{3}^{\mathrm{coll}}<U_{3}^{\mathrm{crit}}$, as
we showed recently~\cite{carr2004b}.  The critical nonlinearity
for collapse corresponds to an eigenvalue $\mu ^{\mathrm{coll}}$,
while $U_{3}^{\mathrm{crit}}$ corresponds to $\mu =0$. Therefore a
simple way to state this condition is
\begin{equation}
\mu ^{\mathrm{coll}}<0\,.
\end{equation}
In Fig.~\ref{fig:2}(a),  $\mu ^{\mathrm{coll}}$ is shown as a
function of the trap parameter $\alpha $ for $V_{0}=0$. Clearly,
in order to observe the transition from a quasi-bound to a bound
state in three dimensions, one needs a finite offset $V_{0}<0$, as
was shown in Ref.~\cite{carr2004b}. For $\alpha \ll 1$, the effect
of a nonzero offset $V_{0}$ on $\mu ^{\mathrm{coll}}$ is
essentially linear: $\mu ^{\mathrm{coll}}\simeq
\mu^{\mathrm{coll}}(V_{0}=0)+V_{0}$. For instance, to observe the
transition from the quasi-bound state to the bound one, it is
required that $V_{0}\leq -0.273$ for $\alpha =1/4$. For larger
$\alpha $, the relation between $\mu $ and $V_{0}$ becomes
nonlinear. For example, in the case of $\alpha =1$, the necessary
condition is $V_{0}\leq -1.495$. Since such values of $\alpha $
are experimentally irrelevant, we do not illustrate this regime.

Inspection of Fig.~\ref{fig:2}(a) reveals that the curve
terminates at $\alpha =0.2$. For $\alpha >0.2$ and $V_{0}=0$,
there is no quasi-bound state at all. Another way of stating this
is that for any $\alpha $, there is a critical
$V_{0}^{\mathrm{crit}}$ such that
\begin{equation}
V_{0}\leq V_{0}^{\mathrm{crit}}
\end{equation}is required to obtain any stationary solution of the form described by the
ansatz of Eq.~(\ref{eqn:ansatz}). A plot expressing this condition
is displayed in Fig.~\ref{fig:2}(b). This is an important
consideration in choosing the correct experimental regime to
observe quasi-bound states in three dimensions.

\begin{figure}[t]
\centerline{
\includegraphics[width=3in,angle=0,keepaspectratio]{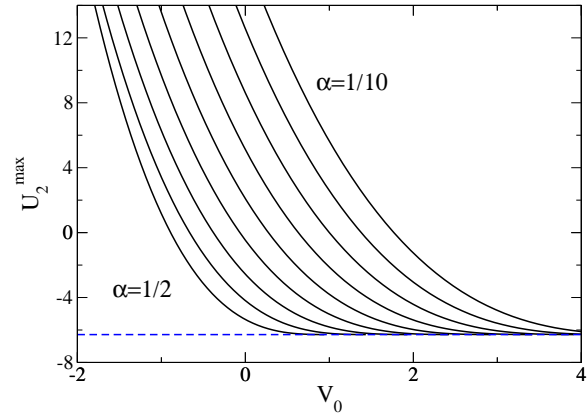}}
\caption{\textit{Ground state parameters, two dimensions.} (a)
Shown is the dependence of the maximum nonlinearity
$U_2^{\mathrm{max}}$ to obtain a quasi-bound state on the
potential offset $V_0$. The curves show
$\protect\alpha=1/2,1/3,\ldots,1/10$ from bottom left to top
right. The dashed blue line shows the collapse point,
$U_2^{\mathrm{coll}}=-2\protect\pi$. In this harmonic times
Gaussian potential, one is therefore limited to the range of
nonlinearities $U_2^{\mathrm{coll}}<U_2<U_2^{\mathrm{max}}$. Note
that for $U_2\leq U_2^{\mathrm{coll}}$ there is no stationary
state at all.} \label{fig:3}
\end{figure}

In 1D there is no collapse at all. In 2D and in free space,
initial conditions lead to collapse for
$U_{2}<U_{2}^{\mathrm{coll}}$ and expansion for
$U_{2}>U_{2}^{\mathrm{coll}}$, while for
$U_{2}=U_{2}^{\mathrm{coll}}$ an unstable stationary state is
obtained, known as the Townes soliton~\cite{sulem1999}. This weak
collapse is different from the strong collapse which occurs in
3D~\cite{sulem1999}. An appropriate external potential can
stabilize the expanding regime, producing a stable solution,
rather than a metastable one as in 3D (see, for instance, the
discussion in Ref.~\cite{carr2004k}). The value
$U_{2}^{\mathrm{coll}}$ which determines the collapse threshold
can be derived from Eqs.~(\ref{eqn:mu1})-(\ref{eqn:U01}) by taking
$\rho \rightarrow 0^{+}$. Then $\mu \rightarrow -\infty $ and
\begin{equation}
U_{2}^{\mathrm{coll}}=-2\pi \,.
\end{equation}
A third critical nonlinearity is given by the point at which the
eigenvalue $\mu =\max [V(r)]$, i.e., where the quasi-bound state
ceases to exist. We called this maximum nonlinearity
$U_{2}^{\mathrm{max}}$. The range of allowed nonlinearities for
obtaining a stationary state is therefore given by
$U_{2}^{\mathrm{coll}}<U_{2}<U_{2}^{\mathrm{max}}$. In this range
all solutions are stable or quasi-stable. Outside of this range
there are no stationary states. Figure~\ref{fig:3} illustrates the
dependence of $U_{2}^{\mathrm{max}}$ on the other parameters in
the problem. The collapse point is shown as a dashed blue line.
Note that the inequalities
$U_{2}^{\mathrm{coll}}<U_{2}^{\mathrm{crit}}<U_{2}^{\mathrm{max}}$
hold for all shapes of the trap, unlike the 3D case.

A technical point concerning two dimensions is that there is always an
additional solution with large $\rho$ which is radially unstable within the
context of our Gaussian ansatz and the VK criterion. This corresponds to a
ring of condensate around the trap. Its variational form and stability can
be more appropriately investigated with a vortex-like ansatz, as we briefly
discuss in Sec.~\ref{sec:vortex}.

A clear experimental signature for a quasi-bound to bound state
transition is a change in the lifetime of the condensate. Suppose
that the lifetime due to tunnelling of the wave function is on the
order of or much less than that imposed by three body processes,
scattering with the background gas in the vacuum, and other
sources of loss. One can then determine the difference between a
quasi-bound and bound state by measuring the number of atoms in
the trap as a function of time. It is necessary to choose the
right experimental parameters in order to have a realistically
observable lifetime. For instance, typical BEC lifetimes are on
the order of ten to a hundred seconds. However, the mean-field
induced dynamical time scales are typically on the order of
milliseconds. Therefore, a lifetime due to tunnelling on the order
of 10 milliseconds to 1 second is desirable.

\begin{figure}[t]
\centerline{
\includegraphics[width=3in,angle=0,keepaspectratio]{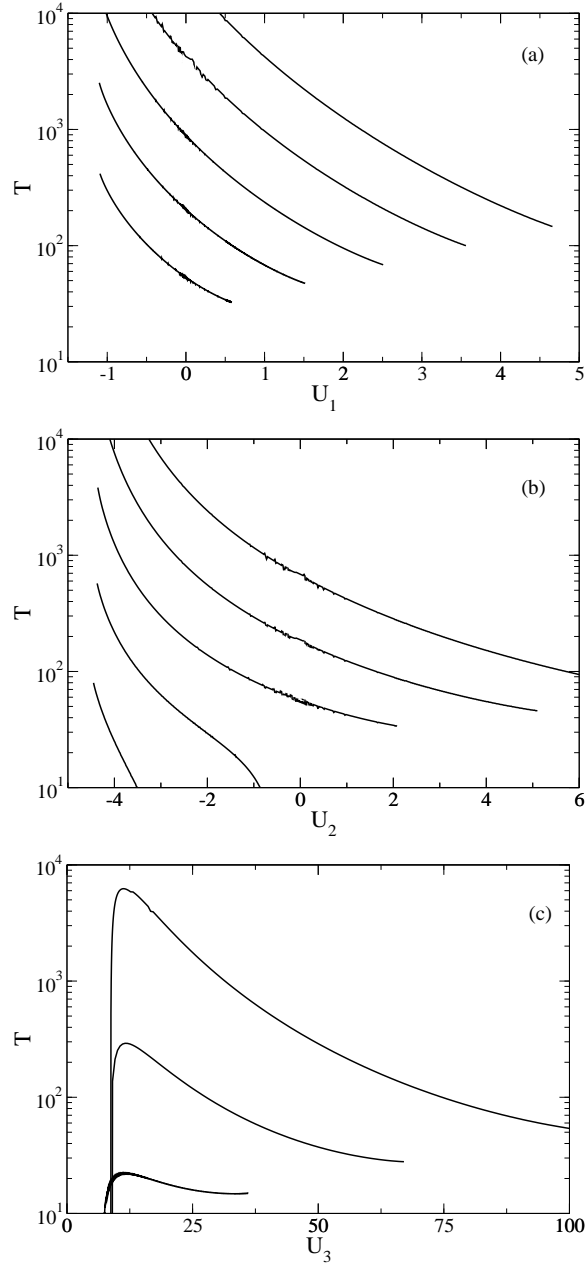}}
\caption{\textit{Ground state lifetime.} Shown is the lifetime $T$
of the BEC, scaled to the trap frequency $\protect\omega$, for
zero potential offset $V_{0}$ and for several values of the
trap-shape parameter $\protect\alpha \equiv \ell
_{\mathrm{ho}}^{2}/\ell _{\mathrm{Gauss}}^{2}$, as a function of
the initial nonlinearity $U_{D}\propto N$. (a) $D=1$, $V_{0}=0$,
$\protect\alpha =1/4,1/5,1/6,1/7,1/8$ (from bottom to top). (b)
$D=2$, $V_{0}=0$, $\protect\alpha =1/4,1/5,1/6,1/7,1/8$ (from
bottom to top). (c) $D=3$, $V_{0}=-2$, $\protect\alpha
=1/4,1/6,1/8$ (from bottom to top). The leftmost point of each
curve approaches the value $U_{D}^{\mathrm{crit}}$, where
$\protect\mu \rightarrow 0^{+}$, the quasi-bound state disappears,
and the lifetime $T\rightarrow \infty $. The rightmost point
corresponds to $U_{D}^{\mathrm{max}}$, where a quasi-bound state
is no longer supported by the potential.} \label{fig:4}
\end{figure}

In Fig.~\ref{fig:4} we show the lifetime for several values of
$\alpha $. Panels (a)-(c) pertain to one, two, and three
dimensions. The lifetime is calculated from
Eq.~(\ref{eqn:lifetime}), as discussed in Sec.~\ref{sec:setup}.
The leftmost endpoint of the curves corresponds to
$U_{D}^{\mathrm{crit}}$. When $U_{D}^{\mathrm{crit}}>0$ the final
state is bound and has a nonzero number of atoms. Three examples
are shown for $D=3$ in Fig.~\ref{fig:4}(c). The rightmost endpoint
of the curves corresponds to $U_{D}^{\mathrm{max}}$. When
$U_{D}^{\mathrm{max}}<0$, the final state is unbound and the
remaining atoms spill out over the top of the trap without
tunnelling. Two examples are shown for $D=2$ in
Fig.~\ref{fig:4}(b).

Since the lifetime is scaled to the trap frequency $\omega$, one can convert
the range of $T$ to milliseconds easily. For instance, for a trap of angular
frequency $\omega=2\pi\times 100$ Hz, the range $T=10$ to $10^4$ shown in
Fig.~\ref{fig:4} corresponds to 16 ms to 16 s. Thus the trap parameters
chosen for the figure result in experimentally observable lifetimes.

The relationship between the tunnelling rate, defined by
Eq.~(\ref{eqn:rate}), and the lifetime, defined by
Eq.~(\ref{eqn:lifetime}), can be very non-intuitive with respect
to what is known from single-particle tunnelling in the linear
Schr\"{o}dinger equation. For example, in Fig.~\ref{fig:4}(c) the
lifetime approaches zero on the left hand side as
$U_{3}\rightarrow U_{3}^{\mathrm{crit}}>0$.  This can be
physically understood in terms of the tunnelling of the
wavefunction through the effective potential.  As the real part of
the chemical potential approaches zero from above, so that the
quasi-bound state becomes a bound state, the attractive
nonlinearity pulls down the potential barrier.  That is, the
effective potential barrier shrinks.  In the case of three
dimensions, and for $\mathrm{Re}(\mu)<0$, it is well-known that
the attractive nonlinearity can dominate over both the kinetic
energy and the potential energy, leading to strong self-focusing
and collapse. Here, one observes that the attractive nonlinearity
already dominates over the potential energy barrier for
$\mathrm{Re}(\mu)
> 0$. This causes the \emph{effective} potential barrier
to disappear for a very narrow region of $U_3$ near the formation
of the bound state, so that the lifetime goes to zero.  This
region is so narrow that it is not experimentally relevant, since
number fluctuations cannot be finely controlled in BEC
experiments, and $U_3 \propto N$.

One can more rigorously understand the limiting values of the
lifetime in the form of the power law of the tunnelling rate
$\gamma $ as $U_{3}\rightarrow \left( U_{3}^{\mathrm{crit}}\right)
^{+}$. Suppose the power law takes the form
\begin{equation}
\gamma (U_{D})\simeq \left( U_{D}-U_{D}^{\mathrm{final}}\right) ^{p}\,,
\end{equation}where $p$ is a real positive constant and $U_{D}^{\mathrm{final}}$
can be zero, $U_{D}^{\mathrm{max}}$, or $U_{D}^{\mathrm{crit}}$,
depending on the allowed domain of quasi-bound states (see the
discussion of the appropriate integration limits following
Eq.~(\ref{eqn:lifetime})). By expanding the integral according to
$\int_{a}^{a+\epsilon }dx\,f(x)\simeq \epsilon f(a)$, one finds
that for $0\leq p<1$ the lifetime approaches zero, despite the
rate approaching zero. For $p>1$ the lifetime approaches infinity,
while for $p=1$ it approaches a nonzero constant. We find
numerically for the ground state that whenever
$U_{D}^{\mathrm{final}}=U_{D}^{\mathrm{crit}}>0$ or
$U_{D}^{\mathrm{final}}=U_{D}^{\mathrm{max}}<0$ the power law of
the rate is such that the lifetime approaches zero at these
critical points. For $U_{D}^{\mathrm{final}}=0$, which occurs when
$U_{D}^{\mathrm{crit}}<0$ and $U_{D}^{\mathrm{max}}>0$, $p=1$ and
the lifetime approaches a nonzero constant.

Finally, we note one other important point about the lifetime. For
$U_D=0$, where there is no mean field, one recovers the correct
single-particle $T=1/\gamma$, as can also be seen by consideration
of Eq.~(\ref{eqn:lifetime}).  This is a limiting case in the
center of a smooth lifetime curve in Fig.~\ref{fig:4} (see also
Figs.~\ref{fig:6} and~\ref{fig:9} below). Therefore, it strongly
supports the validity of the mean field approximation.

\section{Tunnelling of Excited States}

\subsection{Soliton Tunnelling in One Dimension}

\label{sec:soliton}

\begin{figure}[t]
\centerline{
\includegraphics[width=3in,angle=0,keepaspectratio]{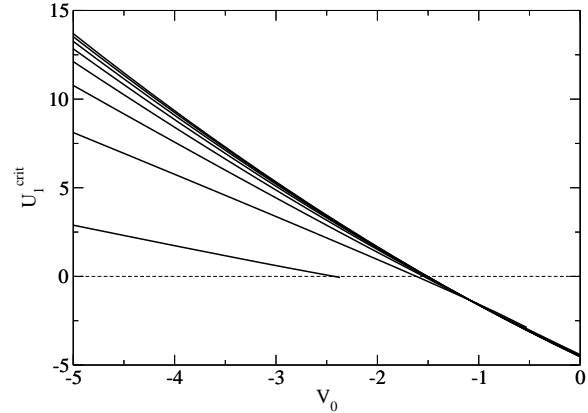}}
\caption{\textit{Excited state transition, soliton.} The critical
nonlinearity for a transition from a quasibound to a bound state
is shown, in analogy to Fig.~\protect\ref{fig:3}(a). Positive
values of $U_{1}^{\mathrm{crit}}$ pertain to the dark soliton;
negative values are for a bright twisted soliton. The dashed line
divides the two regimes. The curves correspond to increasing
values of $\protect\alpha =\ell _{\mathrm{ho}}^{2}/\ell
_{\mathrm{Gauss}}^{2}$: $\protect\alpha =1,1/4,1/9,\ldots $ from
bottom to top of the $y$-axis intercept. For sufficiently small
$\protect\alpha $, all curves lie on the final dark line.}
\label{fig:5}
\end{figure}

The variational ansatz
\begin{equation}
\Psi _{\mathrm{sol}}=A\,r\exp \left( -\frac{r^{2}}{2\rho ^{2}}\right)
\label{eqn:vortexansatz}
\end{equation}models the excited state with a single node in one dimension.  This
is a \emph{dark soliton} for $U_{1}>0$, \textit{i.e.}, a repulsive
BEC in 1D, and an antisymmetric, or \emph{bright twisted} soliton
for $U_{1}<0$, \textit{i.e.}, an attractive BEC in 1D. The
variational equations for the chemical potential and nonlinearity
may be derived by means of the methods outlined in
Sec.~\ref{sec:setup}:
\begin{eqnarray}
\mathrm{Re}(\mu ) &=&\frac{1}{4}\left\{ -9\rho ^{-2}+(1+\alpha \rho
^{2})^{-7/2}\left[ 4V_{0}+(15-16V_{0}\alpha )\rho ^{2}\right. \right.
\nonumber \\
&&\left. \left. -5\alpha (3+4V_{0}\alpha )\rho ^{4}\right] \right\}
\,,\,\,\,\,\,\,  \label{eqn:mu2} \\
U_{1} &=&\frac{2\sqrt{2\pi }}{\rho }\left\{ -2+(1+\alpha \rho
^{2})^{-7/2}\rho ^{4}\left[ 2-4V_{0}\alpha \right. \right.   \nonumber \\
&&\left. \left. -\alpha (3+4V_{0}\alpha )\rho ^{2}\right] \right\} \,,
\label{eqn:U02}
\end{eqnarray}
A plot of $U_{1}^{\mathrm{crit}}$, the dependence of the critical
nonlinearity for the transition from a quasi-bound state to a
bound state, is shown in Fig.~\ref{fig:5}. The transition point is
only very weakly dependent on $\alpha $ for $\alpha \ll 1$. This
regime is an experimental relevant one.

Figure~\ref{fig:6} shows the main experimental observable
characteristic of quasi-bound solitons, namely, the lifetime. The
node causes the wavefunction to become somewhat broader than the
ground state. This requires smaller values of $\alpha $ in order
to provide sufficiently thick walls to hold the quasi-bound
condensate, in comparison to Fig.~\ref{fig:4}(a). As was true for
the ground state, the leftmost endpoint of the curves corresponds
to $U_{1}^{\mathrm{crit}}$, while the rightmost endpoint
corresponds to $U_{1}^{\mathrm{max}}$. For $\alpha =1/6$ and
$V_{0}=0$, one finds $U_{1}^{\mathrm{max}}<0$. Thus, after losing
some atoms via tunnelling through the barrier, the final state
takes the form of two bright solitons with a phase difference of
$\pi $ which travel in opposite directions away from the potential
center.

\subsection{Vortex Tunnelling in Two Dimensions}

\label{sec:vortex}

Under the assumption of an axisymmetric stationary state in two
dimensions, one may take the variational ansatz for a vortex state
as
\begin{equation}
\phi _{m}^{\mathrm{vor}}=A\,r^{m}\exp \left( -\frac{r^{2}}{2\rho
^{2}}\right) \exp (im\theta )\,,
\end{equation}where $m$ is the winding number, or topological charge. Then,
transforming the ansatz according to Eq.~(\ref{eqn:2dtransform})
and minimizing the Lagrangian with respect to
Eq.~(\ref{eqn:gpe2d}) with effective
potential~(\ref{eqn:effpot2d}), one derives the following system
of equations for the chemical potential:
\begin{eqnarray}
\mathrm{Re}(\mu ) &=&\frac{1}{2\,\Gamma (1+m)}\left\{ -\frac{\Gamma (2+m)}
{\rho ^{2}}+(1+\alpha \rho ^{2})^{-3-m}\right.   \nonumber \\
&&\left. \times \left[ -2V_{0}(1+\alpha \rho ^{2})(-1+(1+2m)\alpha \rho
^{2})\Gamma (1+m)\right. \right.   \nonumber \\
&&\left. \left. -\rho ^{2}(-3+(1+2m)\alpha \rho ^{2})\Gamma (2+m)\right]
\right\} \,,\,\,\,\,\,\,  \label{eqn:mu3} \\
U_{2} &=&-\frac{\Gamma (1+m)\Gamma (2+m)\,2^{1+2m}\pi }{\Gamma (1+2m)}
\left\{ 1+(1+\alpha \rho ^{2})^{-3-m}\right.   \nonumber \\
&&\left. \times \rho ^{4}\left[ -1+2V_{0}\alpha +\alpha (1+m+2V_{0}\alpha
)\rho ^{2}\right] \right\} \,,  \label{eqn:U03}
\end{eqnarray}As a consistency check, note that, for the case $m=0$, one recovers the same
result as given by Eqs.~(\ref{eqn:mu1}) and~(\ref{eqn:U01}) with
$D=2$.

\begin{figure}[t]
\centerline{
\includegraphics[width=3in,angle=0,keepaspectratio]{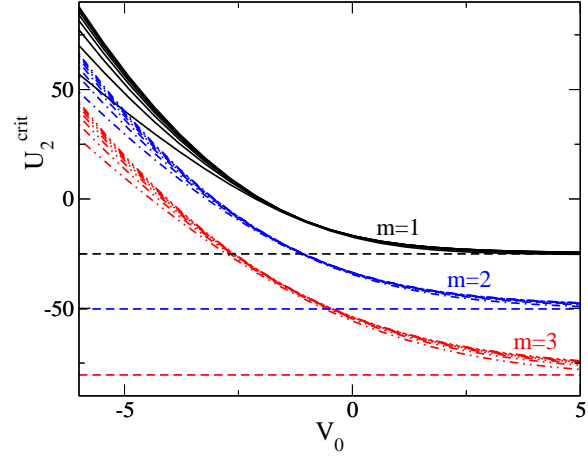}}
\caption{(color online) \textit{Excited state transition, vortex.}
The critical nonlinearity for a transition from a quasibound to a
bound state is shown, in analogy to Fig.~\protect\ref{fig:3}(b),
for values of the winding number $m=1$ (solid black curves), $m=2$
(dot-dashed blue curves), and $m=3$ (dot-dot-dashed red curves).
Each set of curves rises along the $y$-intercept for decreasing
$\protect\alpha $ as $\protect\alpha =3^{-2},\,4^{-2},\,\ldots
,\,10^{-2}$. The dashed horizontal lines show the nonlinearities
for which the solutions collapse:
$U_{2}^{\mathrm{coll}}=-8\protect\pi ,-16\protect\pi
,-(128/5)\protect\pi $ for $m=1,2,3$. Note that
$U_{2}^{\mathrm{crit}}>U_{2}^{\mathrm{coll}}$ for all
$V_{0},\protect\alpha ,m$.} \label{fig:7}
\end{figure}

\begin{figure}[t]
\centerline{
\includegraphics[width=3in,angle=0,keepaspectratio]{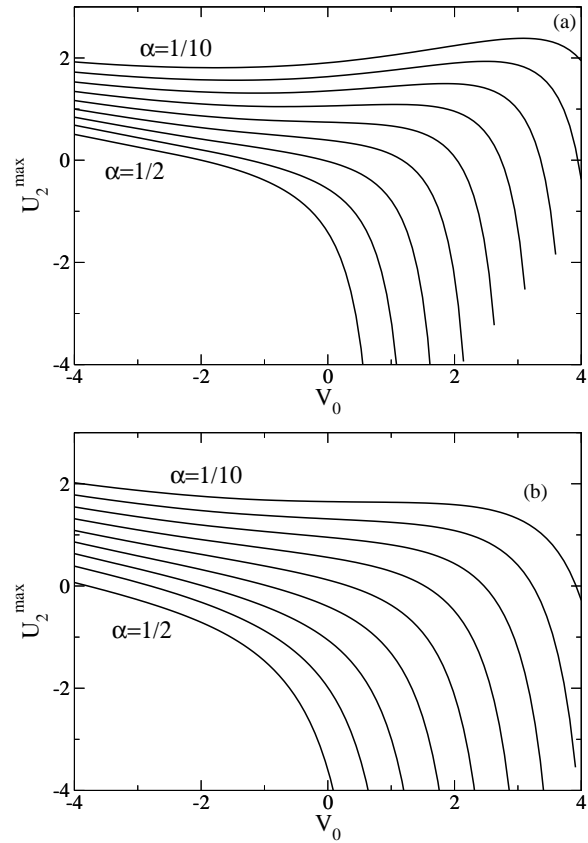}}
\caption{\textit{Excited state parameters, vortex.} Shown is the
dependence of the maximum nonlinearity for confinement of a
vortex, $U_{2}^{\mathrm{max}}$, on the parameters
$V_{0},\protect\alpha ,m$. Here $V_{0}$ is the potential offset,
$\protect\alpha $ the trap-shape parameter, and $m $ the vortex
winding number. Panels (a) and (b) illustrate $m=1$ and $2$,
respectively, while $\protect\alpha =1/2,1/3,\ldots ,1/10$ from
bottom to top in each panel. The asymptotic right hand value of
each curve represents the critical parameter set for which a
vortex is pushed outside the potential, i.e., where it becomes
pinned rather than confined.} \label{fig:8}
\end{figure}

The collapse point may be obtained by expansion of
Eqs.~(\ref{eqn:mu3}) and~(\ref{eqn:U03}) in small $\rho $, in the
same way as was done in Sec.~\ref{sec:ground}. One thus finds
\begin{equation}
U_{2}^{\mathrm{coll}}=-\frac{2\pi ^{3/2}\Gamma (2+m)}{\Gamma (1/2+m)}\,.
\end{equation}For $m=0,1,2,3$ one obtains $U_{2}^{\mathrm{coll}}=-2\pi ,\,-8\pi
,\,-16\pi ,\,-(128/5)\pi $. The presence of the vortex decreases
the collapse point, as was pointed out some time
ago~\cite{ruprecht1995,dodd1996,alexander2002}. The critical point
for a transition from a quasi-bound to a bound state is shown in
Fig.~\ref{fig:7} for $m=1,2,3$. Like the ground state in two
dimensions, the nonlinearity is constrained to the range
$U_{2}^{\mathrm{coll}}<U_{2}<U_{2}^{\mathrm{max}}$. In
Fig.~\ref{fig:8} the allowed range is shown as a function of
$\alpha $ and $V_{0}$ for $m=1,2$.

\begin{figure}[t]
\centerline{
\includegraphics[width=3in,angle=0,keepaspectratio]{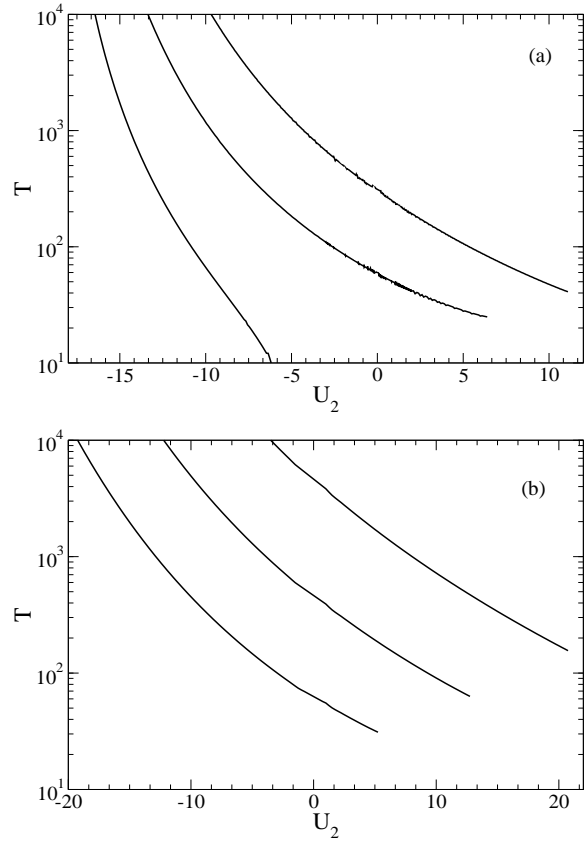}}
\caption{\textit{Excited state lifetime, vortex.} Shown is the
lifetime of the condensate with a vortex in two dimensions. (a)
Winding number $m=1$ and trap-shape parameter $\protect\alpha
=1/8,\,1/10,\,1/12$. (b) $m=2$ and $\protect\alpha
=1/14,/,1/16,\,1/18$. The leftmost point of each curve, which is
not shown on the plot, approaches $U_{2}^{\mathrm{crit}}$, where
$\protect\mu \rightarrow 0^{+}$, the quasi-bound state disappears,
and the lifetime $T\rightarrow \infty $. The rightmost point
corresponds to $U_{2}^{\mathrm{max}}$, where a quasi-bound state
is no longer supported by the potential (see
Fig.~\protect\ref{fig:8}).} \label{fig:9}
\end{figure}

The lifetime of the condensate with a vortex is shown in
Fig.~\ref{fig:9} for $m=1,2$. For higher winding numbers the
ansatz becomes broader, and it is necessary to choose smaller
$\alpha $ in order to contain the vortex. This is reflected in the
range of $\alpha $ in Fig.~\ref{fig:9}, $\alpha =1/8$ to $1/12$
for $m=1$ and $\alpha =1/14$ to $1/18$ for $m=2$. Note that,
unlike for the cases illustrated in Figs.~\ref{fig:4}
and~\ref{fig:6}, here $T\rightarrow \infty $ as $U_{2}\rightarrow
U_{2}^{\mathrm{crit}}$. In Fig.~\ref{fig:9}(a), for $\alpha =1/8$
the atoms first tunnel through the barrier. When $U_{2}$ reaches
$U_{2}^{\mathrm{max}}<0$, the condensate spills out over the top
of the potential barrier. Then it expands towards $r=\infty $ as a
bright ring soliton~\cite{firth1997,carr2004k}, pushed outwards by
the Gaussian tail of the potential. At a critical value of the
ring radius, it undergoes azimuthal modulational
instability~\cite{hasegawa1980}.

We have not considered azimuthal instabilities.  These should be
manifest for nonzero winding number in the attractive case in 2D
in certain parameter regimes~\cite{firth1997,saito2004}. A simple
stability criterion is that the wavelength of modulational
instability be longer than the ring circumference $2\pi \rho $
(for details, see~\cite{carr2004k} and references therein). A more
sophisticated treatment via a full linear stability analysis
(using the Bogoliubov equations) is left for a future work.

We note that for nonzero winding number there is an additional
stationary state which has a radius greater than that of the
potential peak. For repulsive nonlinearity, this is a radially
unstable pinned vortex with an initial core size artificially
controlled via the trapping potential. For attractive
nonlinearity, this state can be radially stabilized for
sufficiently strong nonlinearity, as may be shown via the VK
criterion of Eq.~(\ref{eqn:vk}). However, in this regime we expect
the azimuthal instability to dominate, so we have not presented it
here in any detail.

\section{Conclusion}

\label{sec:conclusion}

In this study we suggested a straightforward macroscopic quantum
tunnelling experiment which requires little or no modification of
existing experimental apparatus. We assumed a trap consisting of a
parabolic potential times a Gaussian envelope, which models
typical optical traps used in experiments. Such a trap supports
both bound and quasi-bound states. Using a variational-WKB
mean-field formalism, we calculated four experimental observables:
the lifetime $T$ of quasi-bound condensates; the maximum
nonlinearity $U_{D}^{\mathrm{max}}$ for which a quasi-bound state
exists; the critical nonlinearity $U_{D}^{\mathrm{crit}}$ for
which a quasi-bound state becomes a bound state; and, for two and
three dimensions, the collapse nonlinearity
$U_{D}^{\mathrm{coll}}<0$.

By adjusting the initial number of atoms and/or the atomic
interaction strength, as may be achieved via a Feshbach
resonance~\cite{vogels1997,inouye1998,timmermans1999}, we showed
how a quasi-bound condensate can be adiabatically transformed into
a bound one, and vice versa. We presented in detail the parameter
regimes in which such a transformation can be performed for both
repulsive and attractive condensates.  In two dimensions, we found
that the relations
$U_{2}^{\mathrm{max}}>U_{2}^{\mathrm{crit}}>U_{2}^{\mathrm{coll}}$
always hold; in three dimensions, it is necessary to add a
negative offset to the potential at $r=0 $ in order to observe the
quasi-bound to bound transition without provoking collapse of the
condensate. This offset can be constructed with an off-resonant
blue- or red-detuned laser focused at the origin of the trap.

We found that, for reasonable experimental parameters, one could
observe the tunnelling of a BEC through the walls of an optical
trap on time scales of 10 milliseconds to 10 seconds. We showed
that tunnelling can lead to a final state which is quite different
from that obtained via the linear Schr\"odinger equation. For
$U_D^{\mathrm{crit}}>0$, an initially quasi-bound repulsive
condensate approaches a bound state with a finite number of atoms
remaining in the well. For $U_D^{\mathrm{max}}<0$ an initially
quasi-bound attractive condensate approaches an unbound state for
which the atoms spill out over the top of the well. This is
observable as a pair of counter-propagating bright solitons in one
dimension, and as an azimuthally unstable bright ring soliton in
two dimensions~\cite{firth1997,carr2004k}.

We showed previously that our variational-WKB method accurately
determines the tunnelling rate and critical points at the level of
1\% or better, except near the collapse points, where it is
accurate to within about 10\%~\cite{carr2004b}. It is also
possible to test our method via exact solution with a small number
of atoms. For example, it has been shown that mean field effects
are strongly evident in as few as three bosons confined in an
external potential~\cite{blume2002}.

Finally, we note that there is a direct analogy between the
operation of a laser and the tunnelling of a BEC through the walls
of an optical trap.  In a laser operating well above threshold,
the coherent state in the cavity is robust against removal of
single photons via tunnelling through a thin barrier.  This is the
essence of a coherent state.  The mean field of the Bose
condensate in an optical trap is phase-coherent and robust in the
same way.  Therefore we expect our mean field approximation to be
an excellent one for a large initial number of atoms in the trap.
When the initial number of atoms becomes very small, the
condensate is liable to phase fluctuations.  However, as our
lifetime definition and figures have the right single-particle
limit for zero mean field, we expect that the mean field theory
also gives good results in this region.

We acknowledge useful discussions with Yehuda Band, Nimrod
Moiseyev, and Brian Seaman. L. D. Carr and M. J. Holland
acknowledge the support of the U.S. Department of Energy, Office
of Basic Energy Sciences via the Chemical Sciences, Geosciences
and Biosciences Division. B. A. Malomed acknowledges support from
the Israel Science Foundation, through grant No. 8006/03, and
appreciates the hospitality of JILA (University of Colorado,
Boulder).

\begin{figure}[t]
\centerline{
\includegraphics[width=3in,angle=0,keepaspectratio]{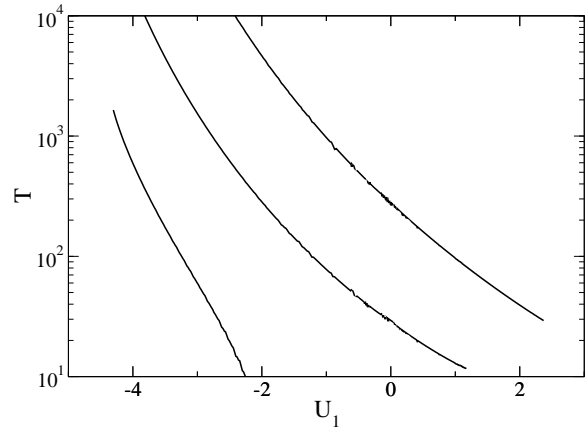}}
\caption{\textit{Excited state lifetime, soliton.} Shown is the lifetime of
the excited state with a single node in one dimension, i.e., a dark soliton
for $U_{1}>0$ and a bright twisted soliton for $U_{1}<0$. The parameters are
$V_{0}=0$ and $\protect\alpha =1/6,1/8,1/10$ from left to right.}
\label{fig:6}
\end{figure}

\end{document}